# A Novel architecture for improving performance under virtualized environments

A.P. Nirmala  
Research Scholar, Karpagam University, Coimbatore &  
Assistant Professor, New Horizon College of Engg.,  
Bangalore, India  
nirmala_ap@yahoo.com

Dr. R. Sridaran  
Dean, Faculty of Computer Applications,  
Marwadi Education Foundation's Group of Institutions,  
Rajkot, India  
sridaran.rajagopal@gmail.com

*Abstract*—**Even though virtualization provides a lot of advantages in cloud computing, it does not provide effective performance isolation between the virtualization machines. In other words, the performance may get affected due the interferences caused by co-virtual machines. This can be achieved by the proper management of resource allocations between the Virtual Machines running simultaneously. This paper aims at providing a proposed novel architecture that is based on Fast Genetic K-means++ algorithm and test results show positive improvements in terms of performance** *improvements over a similar existing approach*.

*Keywords*- Virtualization, Performance, Performance Interference, Scheduling Algorithm, Throughput

## I. INTRODUCTION

In the recent era, virtualization technology provides the advantages in the form of better manageability, optimistic provisioning and minimizing the cost in current cloud computing environments.

Virtualization allows sharing of server resources on-demand thereby creating new business opportunities. This leads to developments of new delivery models for a wider set of enterprise services. Thus, virtualization is a key enabling factor not only for Cloud Computing but also for utility computing paradigm [2][16]. However, virtualization may also lead to the contention of shared resources on each platform between virtual machines (VMs) involved which needs to be addressed.

Virtualization technology enables diverse applications to run in the isolated environments by creating multiple VMs on a single physical machine and managing resource sharing across VMs by virtual machine monitor (VMM) technology.

VMMs or hypervisors from VMware™, Xen™ community, Microsoft™ and others manage the VMs running on a single platform and ensure that they are functionally isolated from one another as shown in Fig. 1.1 [5][16].

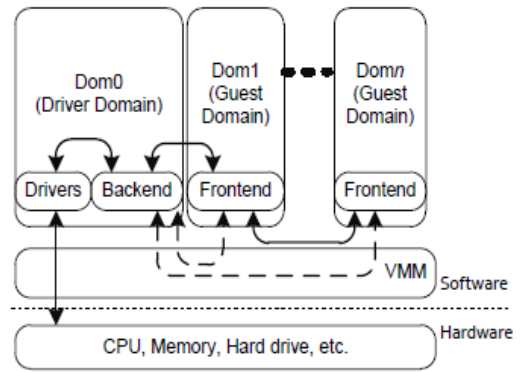

Fig. 1.1 Virtualized cloud environment.

The VMM is responsible for allocating basic resources such as CPU cycles, memory capacity, disk and network I/O bandwidth. At a high level, allocating a specific share of physical resources to a VM results in a specific performance that is measurable using certain performance metrics such as response time and/or throughput. However, as all VMs share the same physical resources, they also mutually influence each other's performance. According to X. Pu et al. [5], VMMs (hypervisors) have the abilities to slice down resources and allocate the shares to different VMs where the applications running on one VM may still affect the performance of applications running on its neighborhood VMs.

However virtualization provides features such as security, fault and environment isolations, it does not help in offering performance isolation between VMs in an effective manner. Y. Koh et al. [3], describes that a user running the same VM that belongs to the hardware but at different times will realize wide disparity in performance based on the work carried out on other VMs on that physical host.

It is essential to develop architectural techniques that ensure appropriate sharing of resources allocated to VMs running simultaneously based on their importance or their behavior. Effective management of virtualized cloud environments introduces new and unique challenges, such as efficient CPU scheduling for VMs, effective allocation of VMs to handle both CPU intensive and I/O intensive workloads. Based on this, many novel scheduling algorithms can be thought of. These algorithms may have their primary objectives as either of minimizing the negative impacts of co-located applications or improving the overall system performance [4][1]. Ron C. Chiang et al. in [1] discusses about how the system can make optimized scheduling decisions that lead to significant improvements in both application performance and resource utilization.

According to Ron C. Chiang et al. in [1], K-means algorithm neither guarantees to converge to a global minimum nor to achieve the best optimal solution available. Genetic Algorithm (GA) is used for the purpose of finding the global minima [14]. Ron C. Chiang et al. [1] proposed the K-means++ algorithm to make a good choice of initial k centers to replace K-means algorithm that is designed to pick points that are far away from each other. Instead of choosing the point farthest from chosen points, k-means++ pick each point at random with probability proportional to the squared distance. Thus k-means++ is combined with genetic algorithm to find optimized solution.

This motivates us to propose Fast genetic k-means++, a scheduling algorithm to improve the performance in the virtualized environments. It is implemented by conducting a comprehensive evaluation with a variety of cloud applications to measure the performance in terms of application throughput, runtime and cost.

Fast genetic k-means++ algorithm, aims at improving the performance by scheduling the task to various VMs with minimized interference effects from co-located applications. It helps to reduce the runtime and improve the I/O throughput for data-intensive applications in a virtualized environment. When the task arrives, the scheduler proceeds with generating number of possible assignments of the same based on the incoming tasks and list of available VMs. Then the scheduler makes the decision for scheduling and assigning the task to different servers based on the predictions.

This paper is organized as follows, section 2 provides the related work, section 3 explains the proposed methodology and experimental results are discussed in section 4. Section 5 gives the conclusion with future work..

## II. RELATED WORK

X. Pu et al.. [5] focus on performance interference among VMs running the network workloads in virtualized environments. In their work, system-level characteristics are considered as metrics to identify the impact of running different combinations of workloads of different file sizes on the aggregate throughput. Extensive experiments are conducted to compare and understand better the combination of different workloads and the multiple factors that may cause performance interferences. However, the above work does not describe how to mitigate the I/O interference for data-intensive applications. There are several studies to evaluate the performance degradation of VMs due to interference which are illustrated in [7][3]. Although these studies aim at proposing different types of benchmarks to identify the VM interference but they do not explain how to mitigate the interference effects.

D. Novakovic et al. proposed DeepDive, a system for VM migration [19]. DeepDive identifies interference-inducing VM and shifts VM to destination physical machine on which least interferences are reported. In contrast to DeepDive, in our proposed work, placement of task in appropriate VM alone is considered rather than VM placement itself in order to reduce the interferences.

Recently, Paragon [20], proposed test benchmarks to identify sources of interferences and their impact on co-located applications. Paragon uses previously scheduled applications to identify the best placement for new application with respect to interference in place of profiling. Our proposed work is analogues to Paragon [20] as for as the placement of application is concerned. However, Paragon [20] classifies and schedules the application on the proper hardware platform in a way that minimizes interference rather than scheduling the applications in a VM.

Altino Sampaio et al. have proposed new algorithms to dynamically schedule VMs to minimize the performance interference due to hardware resources such as last-level cache(LLC) sharing [19]. In this approach, the aim is to maximize the rate of completed tasks by constructing performance-efficient computing environments that react to performance degradation arise from sharing of LLC memory. And Q-clouds [18], a QoS-aware control framework developed to mitigate performance interference effects. It uses a multi-input multi-output (MIMO) model that tunes resource allocations to capture the performance interference in a virtualized environment. But, Q-Clouds focuses only on the CPU bound workload. In contrast to the above studies [19], [18], though the cloud applications are data centric, it is essential to address the challenges of the I/O interference when running data-intensive applications in virtualized environments. Thus our work focuses on data-intensive applications.

TRACON framework is proposed by Chiang et al.. [1] with the aim of mitigating the interference effects from co-located data-intensive applications, and thus improving the overall system performance. It is composed of an interference prediction component, an interference-aware scheduler, and task and resource monitors. K-means++ algorithm is implemented in interference-aware scheduler to schedule the task in suitable VMs in a virtualized environment. According to Ron C. Chiang et al. in [1], K-means algorithm neither guarantees to converge to a global minimum nor to achieve the best optimal solution available. Genetic Algorithm (GA) is used for the purpose of finding the global minima [14]. Ron C. Chiang et al. [1] proposed the K-means++ algorithm to make a good choice of initial k centers to replace K-means algorithm that is designed to pick points that are far away from each other. Instead of choosing the point farthest from chosen points, k-means++ pick each point at random with probability proportional to the squared distance. Thus k-means++ is combined with genetic algorithm to find optimized solution. Our proposed work extends this phenomenon by proposing a fast-genetic k-means++ algorithm which is a combination of genetic algorithm with k-means++ to make the optimized decisions in order to improve the overall performance.

### III. PROPOSED WORK

In the proposed work, fast genetic k-means++ is used as a scheduling algorithm to improve the application performance in a virtualized environment. Levenberg-Marquardt method [17], a non-linear model is used to find the optimal solution in predicting the performance. Fast genetic k-means++, a scheduling algorithm is implemented to measure the performance in terms of application throughput, runtime and cost and the progress in the performance results are shown in section 4.

#### A. Terms Used

*1) Interference Prediction Model:* The interference prediction model infers the application performance from the resource consumption observed from multiple VMs. In the proposed work, interference prediction model is constructed using five parameters (controllers) including CPU utilization in VMM, CPU consumption from data processing of application, I/O request, cost and job /cloudlet These parameters are used to read and write throughput, to measure I/O workload from target application in terms of number of requests per second in order to model the CPU consumption from data processing of the application. The parameter CPU utilization is used to increase the accuracy for a virtualized environment.

*a) Non-linear Model:* According to Ron C. Chiang et al. [1], the prediction accuracy in linear model is mostly at par with weighted mean method, thus it cannot be taken as best fit for the observed data. It is essential to opt for an alternative to linear model and weighted mean method leads us to explore the nonlinear models, in particular with the degree of two that is, quadratic models.

The non-linear model is constructed using Levenberg-Marquardt (LM) Method [17]. It is a combination of two minimizing techniques namely gradient descent method and gauss-Newton method. The LM method adaptively varies the parameter updates between the gradient descent update and the Gauss-Newton update,

$$[J^TWJ + \lambda I]h_{im} = J^TW(y - \hat{y}) \quad (1)$$

From the equation, if the parameter $\lambda$ assumes a small value, then it results in a Gauss-Newton update and if a large value then it takes gradient descent update.

After the above update of the relationship, LM algorithm becomes

$$[J^TWJ + \lambda diag(J^TWJ)]h_{im} = J^TW(y - \hat{y}) \quad (2)$$

This algorithm is used to update the parameters in order to obtain optimal solution in virtualized environment. This provides best prediction accuracy for the nonlinear models

$$\hat{N} = c + \sum_{i=1}^{5} \alpha_i^{(1)} . P_{VM\ 1,i} + \sum_{i=1}^{5} \alpha_i^{(2)} . P_{VM\ 2,i} +$$
$$\sum_{i=1}^{5}\sum_{j=1}^{5} \beta_{i,j}^{(1)} . P_{VM\ 1,i} . P_{VM\ 2,j} +$$
$$\sum_{i=1}^{5}\sum_{j=1}^{i-1} \beta_{i,j}^{(2)} . P_{VM\ 1,i} . P_{VM\ 1,j} +$$
$$\sum_{i=1}^{5}\sum_{j=1}^{i-1} \beta_{i,j}^{(3)} . P_{VM\ 2,i} . P_{VM\ 2,j} +$$
$$\sum_{i=1}^{5}\sum_{j=1}^{i-1} \beta_{i,j}^{(4)} . P_{VM\ 2,i} . P_{VM\ 1,j}$$
$$\sum_{i=1}^{5} \gamma_i^{(1)} . P^2_{VM\ 1,i} + \sum_{i=1}^{5} \gamma_i^{(2)} . P^2_{VM\ 2,i} \quad (3)$$

The above equation is given for two VMs VM1 and VM2, each one of them can be assigned with one application. Each model of the proposed system architecture relates to five key parameters for individual VMs, thus resulting in ten variables in both VMs together. N is the response variable representing the run time and $P_{VM\ 1,i}, P_{VM\ 2,i} i \in \{1,2,3,4,5\}$ are the controlled variables representing the application characteristics on VM1 and VM2.

*2) Model training and learning :* Interference profile is generated by running the given application on one VM while the remaining VMs will be executing various workloads in the background where n VMs are involved. This profile has a collection of data on interference effects under different background workloads. This approach supports online learning of interference prediction model that is dynamically modified and monitored for different applications in the cloud platform.

*3) Interference-Aware Scheduling(IAS):* IAS is proposed for scheduling the task to various VMs with minimized interference effects from co-located applications. It aims to reduce the runtime and improve the I/O throughput for data-intensive applications in a virtualized environment. In the proposed work, fast genetic k-means++ algorithm is used for the purpose of improving the performance in the cloud environment. When the task arrives, the scheduler proceeds with generating number of possible assignments of the same based on the incoming tasks and list of available VMs. Then the scheduling process takes place by assigning the task to different servers based on the predictions.

#### B. Fast genetic k-means ++algorithm (FGKA++)

Fig.1 (a) shows the flow of FGKA++ algorithm which starts with the initialization phase, generating the initial task $P_0$. The task in the next generation $P_{i+1}$ is obtained by applying the following genetic operators sequentially: the selection, the mutation and the K-means++ on the current task Pi. The evolution takes place until the termination condition is successfully reached. The algorithm in pseudo code representation is shown in Fig.1 (b)

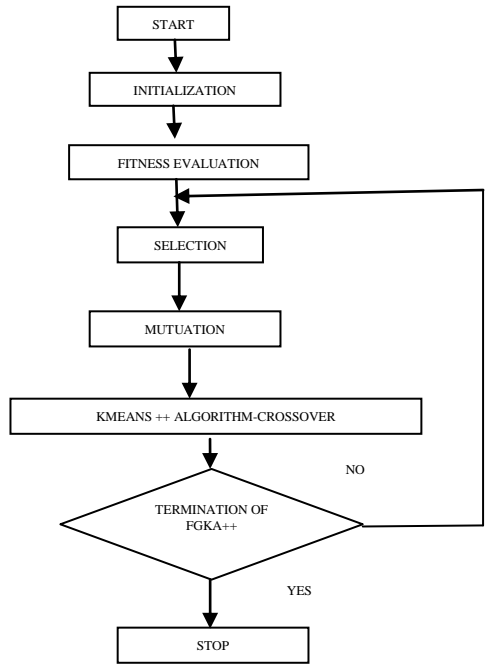

Fig. 1 (a) Flow-chart of FGKA++ algorithm

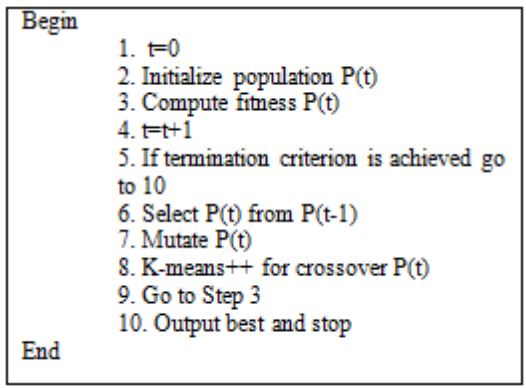

Fig. 1 (b) Pseudo code of FGKA++ algorithm

K-Means algorithm (KMA) provides a method of cluster analysis which aims at portioning of n observations into k clusters. Each of the observation belongs to a cluster with the minimum distance between cluster centre and the observation point. It is done iteratively so that the observation point is at least distance from the centre of cluster. The mean distance between the cluster centre and observation is minimized during this iteration process.

The main limitation with the KMA is that, it neither guarantees to converge to a global minimum nor to achieve the best optimal solution available. Since stochastic optimization approaches are good at avoiding convergence to a local optima, these approaches could be used to find a globally optimal solution. Genetic Algorithm (GA) is used for the purpose of finding the global minima [13].

K-means algorithm is designed to pick points that are far away from each other. To overcome this drawback, Ron C.Chiang et al. [1] proposed the K-means++ algorithm to make a good choice of initial k centers. Instead of choosing the point farthest from chosen points, k-means++ pick each point at random with probability proportional to the squared distance.

Thus k-means++ is combined with genetic algorithm to find optimized solution. The genetic operators used in this approach are the selection, the distance based mutation and the k-means++ operator are explained below.

*1) The Selection Operator:* The task of the next generation is determined by P independent random processes. Each process randomly selects a solution from the current task $\{S_1, S_2, \ldots, S_p\}$ according to the probability distribution, $\{R_1, R_2, \ldots, R_p\}$ defined by

$$R_p = \frac{F(S_p)}{\sum_{p=1}^{p} F(S_p)} \quad (p = 1, \ldots, P) \tag{4}$$

where $F(S_p)$ denotes the fitness value of solution $S_p$.

*2) The Mutation Operator:* The mutation operator performs the function of shaking the algorithm out of a local optimum, and moving it towards the global optimum. During the mutation, $b_n$ is replaced by $b_n'$ for $n = 1, \ldots, N$ simultaneously. $b_n'$ is a cluster number randomly selected from $\{1, \ldots, K\}$ with the probability distribution $\{R_1, R_2, \ldots, R_k\}$ defined by

$$R^k = \frac{1.5 * d\max(X_n) - d(X_n, ck) + 0.5}{\sum_{k=1}^{K}(1.5 * d\max(X_n) - d(X_n, ck) + 0.5)} \tag{5}$$

where $d(X_n, ck)$ is the Euclidean distance between pattern $X_n$ and the centroid $ck$ of the $k^{th}$ cluster, and $d\max(X_n) = \max_K \{d(X_n, ck)\}$.

If the $k^{th}$ cluster is empty, then $d(X_n, ck)$ is defined as 0. The bias 0.5 is introduced to avoid divided by- zero error in the case that all patterns are equal and are assigned to the same cluster in the given solution.

Initially, the above mutation operator ensures that an arbitrary solution, including the global optimum, might be generated by the mutation from the current solution with a positive probability. Second, it encourages that each Xn is moving towards a closer cluster with a higher probability. Third, it promotes the probability of converting an illegal solution to a legal one.

*3) K-Means++ Operator:* In order to speed up the convergence process, one step of the classical K-means++ algorithm, which we call K-means++ operator is introduced. Given a solution that is encoded by b1…bN, we replace bn by bn' for n=1,…,N simultaneously, where bn' is the number of the cluster whose centroid is closest to Xn in Euclidean distance.

## IV. EXPERIMENTAL RESULTS AND DISCUSSIONS

In the proposed work, performance is evaluated for own cloud application. The proposed algorithm is compared with K-means ++ algorithm [1] by measuring the performance based on cost, throughput and execution time. In order to generate a more realistic workload, we randomly choose the datasets, data sizes, and number of processes. The comparison result shows the improvements in the performance of virtualized environment.

### A. Experimental Setup

Fast genetic k-means++ algorithm is executed on different file types such as pdf, image and text files. The computer hardware used for the implementation is of Intel Core2 duo CPU with 3.40 GHz speed, 4GB RAM size and 500 GB hard disk capacity. The software tools consist of Java as programming language in simulator, Mysql as database and Cloudsim for simulating the virtualized environment.

### B. Performance Evaluation

Fast genetic k-means++, scheduling algorithm implemented for scheduling and assigning the tasks to appropriate VMs. When the task arrives, the scheduler proceeds with generating number of possible assignments of the same based on the incoming tasks and list of available VMs. Then the scheduler decides the scheduling and assigning the task to different servers based on the predictions. Thus, the assignment of tasks to suitable VMs with minimum CPU utilization time is experimented on fast genetic k-means++ and k-means++ algorithm and shown the results in Table 1.

Table 1. Scheduling tasks to VMs with CPU utilization on different scheduling algorithms

| Fast Genetic k-means++ algorithm | | | K-means++ algorithm | | |
|---|---|---|---|---|---|
| Job Name | CPU Utilization | VM Id | Job Name | CPU Utilization | VM Id |
| Bp.pdf | 401.43 | 4 | Bp.pdf | 5000 | 2 |
| Phr2.txt | 51.50 | 1 | Phr2.txt | 416.67 | 5 |
| tt.jpg | 54.96 | 6 | tt.jpg | 250 | 1 |

The application throughput is defined as the number of tasks completed in a given period of time. The normalized application throughput is measured for scheduling methods. Fast genetic k-means++ algorithm achieves better throughput compared with k-means++ algorithm is shown in Table 2.

Table 2: Throughput Comparison

| Technique | K-means ++ | Fast Genetic K-means ++ | Improved Throughput |
|---|---|---|---|
| Achieved Throughput (%) | 0.62 | 1.41 | 0.86 |

The Throughput, running cost and the execution time taken by the cloud application are measured and the improvement in the performance is shown in Fig.2 &3.

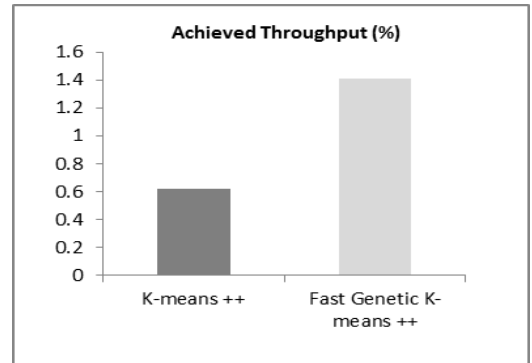

Fig. 2: Throughput Comparison

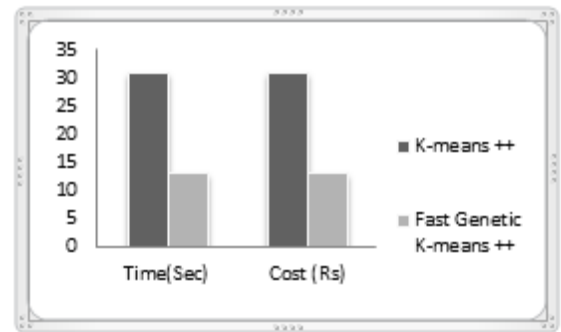

Fig. 3: Comparison of Time and Cost

The cost required for the cloud application is measured based on time utilized in the cloud. The cost of the cloud can be controlled by performing the function with minimal running time. The cost is calculated as. INR 1 per second based on the usage. Hence, the cost for k-means++ is $31 \times 1 = INR\ 31$ whereas, the cost for the fast genetic k-means++ is $13 \times 1 = INR\ 13$. Thus the proposed technique takes lesser time and lower cost while compared with existing technique and the overall cost depends on the cloud providers. Fast genetic k-means++ technique is compared with k-means++ for the same cloud data and the results are shown in Table 3.

Table 3: Comparison of Time and Cost

| Technique | Time(Sec) | Cost (Rs) |
|---|---|---|
| K-means ++ | 31 | 31.00 |
| Fast Genetic K-means ++ | 13 | 13.00 |

## V. CONCLUSION

The proposed architecture shows effective improvements in terms of throughput, time and cost measures. It is believed that the proposed architecture would highly benefit architectures that use simultaneous running VMs without much performance interferences from others. Our future work involves applying the architecture with new age file types and exploring different scheduling algorithms to further improve the performance in the virtualized environment.

REFERENCES

[1]. Ron C. Chiang and H. Howie Huang, "TRACON: Interference-Aware Scheduling for Data-Intensive Applications in Virtualized Environments", IEEE Transactions On Parallel And Distributed Systems, IEEE 2013.

[2]. Mr. Ajey Singh, Dr. Maneesh Shrivastava, "Overview of Security issues in Cloud Computing", International Journal of Advanced Computer Research (IJACR) Volume 2,Number 1,March 2012.

[3]. Y. Koh, R. Knauerhase, P. Brett, M. Bowman, Z. Wen, and C. Pu. An Analysis of Performance Interference Effects in Virtual Environments. IEEE International Symposium on Performance Analysis of Systems and Software (ISPASS), 2007, pp. 200-209.

[4]. Ron C. Chiang and H. Howie Huang. TRACON: Interference-aware scheduling for data-intensive applications in virtualized environments. In Proc. Of SC, pages 1 –12, nov. 2011.

[5]. X. Pu, L. Liu, Y. Mei, S. Sivathanu, Y. Koh, and C. Pu. Understanding performance interference of I/O workload in virtualized cloud environments. In Proc. of CLOUD, pages 51 –58, july 2010.

[6]. A. Gulati, G. Shanmuganathan, I. Ahmad, C. Waldspurger, and M. Uysal. Pesto: online storage performance management in virtualized datacenters. In Proc. of SOCC, pages 19:1–19:14, New York, NY, USA, 2011. ACM.

[7]. H. Shan, K. Antypas, and J. Shalf. Characterizing and predicting the I/O performance of HPC applications using a parameterized synthetic benchmark. In Proc. of SC, pages 42:1–42:12, Piscataway, NJ, USA, 2008. IEEE Press.

[8]. P. Barham, B. Dragovic, K. Fraser, S. Hand, T. Harris, A. Ho, R. Neugebauer, I. Pratt, and A. Warfield, "Xen and the art of virtualization", in Proceedings of the nineteenth
ACM symposium on Operating systems principles, pp. 164-177, 2003.

[9]. C.A. Waldspurger, "Memory resource management in VMware ESX server", ACM SIGOPS Operating Systems Review, Vol. 36, No. si, p. 181, 2002.

[10] B.desLigneris, "Virtualization of Linux based computers: the Linux-VServer project", in 19th International Symposium on High Performance Computing Systems and Applications, pp. 340-346, 2005

[11] S. J. Vaughan-Nichols, "New approach to virtualization is a lightweight", Computer, Vol. 39, No. 11, pp. 12-14, 2006.

[12] D. Arthur and S. Vassilvitskii. k-means++: the advantages of careful seeding. In SODA'07.

[13] K. Krishna and M. NarasimhaMurty, "Genetic K-Means Algorithm", IEEE Transactions On Systems, Man And Cybernetics—Part B:Cybernetics, Vol. 29, No. 3, June 1999.

[14] Yi Lu, Shiyong Lu, Farshad Fotouhi, Youping Deng, Susan J. Brown," FGKA: A Fast Genetic K-means Clustering Algorithm", ACM, 2004.

[15] Rajkumar Buyya, Rajiv Ranjan, Rodrigo N. Calheiros, "Modeling and Simulation of Scalable Cloud Computing Environments and the CloudSim Toolkit: Challenges And Opportunities", in The International Conference on High Performance Computing and Simulation, HPCS2009, pp:1-11.

[16] Ravi Iyer, Ramesh Illikkal, OmeshTickoo, Li Zhao, Padma Apparao, Don Newell, VM3: Measuring, modeling and managing VM shared resources, 2009 Elsevier Computer Networks 53 (2009) 2873–2887.

[17] K. Madsen, H.B. Nielsen, O. Tingleff. Methods for Non-linear Least Squares problems. Informatics and Mathematical Modelling, Technical University of Denmark. April 2004

[18] R. Nathuji, A. Kansal and A. Ghaffarkhah. Q-Clouds : managing performance interference effects for qos-aware clouds. In EuroSys '10.

[19] D. Novakoviíc, NedeljkoVasiíc, Stanko Novakoviíc, DejanKostiíc, and Ricardo Bianchini. DeepDive: Transparently Identifying and Managing Performance Interference in Virtualized Environments. Technical Report 183449, EPFL, 2013

[20] C. Delimitrou et al... Paragon: QoS-aware scheduling for heterogeneous datacenters. In ASPLOS, 2013.

[21] Altino Sampaio, Jorge G. Barbosa, Parallel & Cloud computing, PCC Vol. 2 Iss. 4, 2013PP. 116-125 www.vkingpub.com © 2013 American V-King Scientific Publishing.

AUTHORS PROFILE

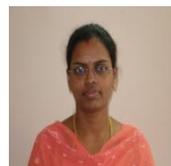

**A. P. Nirmala** received her MCA and M. Phil., degree i*n* Computer Science from Bharathiar University, Coimbatore in 2000 and 2006 respectively. She is pursuing her Ph.D at Karpagam University, Coimbatore, India. She is presently working as Assistant Professor in the Department of Computer Applications, New Horizon College of Engineering, Bangalore, India. She has 12 years of teaching experience. Her research area is Cloud Computing. She has published 1 research paper in an International Journal and presented 6 research papers in National Conferences.

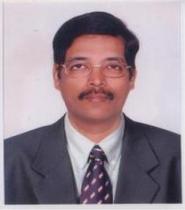**Dr. R. Sridaran** is a Ph.D in Computer Science from Madurai Kamaraj University. He has published 25+ research papers in leading journals and presented in many conferences. He is presently guiding eight research scholars in the areas of Cloud Computing, e-learning and Software Engineering. He has got 20 years of academic experience and served in leading educational institutions at different capacities. He is currently the Dean of Faculty of Computer Applications, Marwadi Education Foundation, Rajkot. He is also the Chairman, Computer Society of India, Rajkot Chapter.